# Excitation of resonant surface plasmons for evanescent waves refocusing by a superlens

Isa Ali

*Department of Physics, Federal University Lokoja, Lokoja, Kogi, Nigeria*

The amplification of evanescent waves by flat superlens requires a near-resonance coupling which has been linked to resonant surface plasmons. A subtle interplay has been proposed to exist between the excitation of well-defined resonant surface plasmons and the focusing capability of a superlens. To gain insights into these resonant modes and their contributions to the amplification and recovery of evanescent waves, we performed simple but robust full-wave FDTD simulations on causal negative index Lorentz models. We found that well-defined pair of resonant surface plasmons is excited whenever a coupling of a diverging transmitted beam and a converging refracted beam occurs at the second interface. The resonant coupling of these modes at the interface led to the excitation of a single interface resonance predicted by the theory. The physical consequence of this resonance is that incident wave energy is pumped into the negative-index material medium and beyond it. These phenomena contribute substantially to the amplification and recovery of the near-fields in the image plane. It is noteworthy that, for evanescent wave refocusing to be achieved in a flat superlens, its thickness, and the source-to-superlens distance should be optimized. This optimization is critical in that an optimal thickness alongside optimal source-to-superlens distance will allow evanescent wave refocusing to dominate material lost. The FDTD simulated result showed an image of the point source inside the superlens and beyond it when its thickness was optimized. The resolution of the image beyond the superlens was $\sim 0.58\lambda$ and this superlens behaves like a near-perfect lens. Overall, these numerically simulated results could serve as useful approximations to the degree of resonant amplification and refocusing of near-fields that can be achieved by flat superlens in near-fields experimental imaging setups.

## I. INTRODUCTION

The anatomy of a superlens [3] is made of a thin silver slab of material with a negative refractive index sandwiched by two positive index media [2]. This lens is capable of reconstructing the image of an object perfectly in the image plane. It is well-known that the sharpness or resolution of conventional lenses is always limited to the wavelength of light however a superlens is capable of overcoming this limit of diffraction by magnifying and recovering evanescent near-fields in the image. This is achieved through the excitation of resonant surface plasmons [2, 3]. This lens is also characterized by permittivity and permeability that are simultaneously negative in a narrow frequency regime such that [1].

Superlens is capable of capturing the propagating far-fields as well as the fast decaying non-propagating near-fields and delivering them in the image plane [2, 4] and it achieves this by applying phase correction to the far-fields and amplifies the near-fields across the slab [2]. This makes it a good candidate when it comes to imaging with super-resolution beyond the diffraction limit and its performance is limited by the effects of absorption in it [2, 4].

It is well-known that conventional lenses, such as concave and convex lenses, focus light whenever there is a contrast in the refractive index and cannot focus light unto an area smaller than a square of the wavelength. However, a superlens is predicted to have the capability of reconstructing all the Flourier components of the source fields including the fast decaying evanescent near-fields through resonant amplification of the fields [5] and refocusing of these fields by surface plasmons [2, 6, 7].

The evanescent near-field components decay exponentially as they propagate away from the source and the primary function of the superlens is to amplify them across it. In the nutshell, the superlens behaves like an amplifier [3]. The near-fields are localized to the immediate vicinity of the source plane and are amplified and recovered with correct amplitudes in the image via surface plasmons resonance [3, 6, 7] or single-interface resonance [9].

One of the fascinating features of the superlens is the appearance of the double-focusing effect [1, 2, 8, 11, 23]. This phenomenon can be imagined when we consider electromagnetic fields radiated by a point source placed in front of superlens. Upon interacting with the first interface of the superlens, they are negatively refracted. This negative refraction effectively reverses propagation direction, and this leads to image formation inside the superlens. The same processes are repeated at the second interface and the subsequent diverging fields are brought to a second focus beyond the superlens.

The superlens is characterized by phase reversal which enables it to refocus the source fields, and it cancels out the acquired phase as the fields propagate away from the source [1, 2]. This lens is capable of compensating for the decay of the evanescent near-fields. These fields decay in amplitudes and not in phase as they propagate from the source. To recover them in the image will require amplification of their components rather than correcting their phases [2]. Thus, a superlens behaves like an amplifier [3]. Pendry [2] demonstrates that electromagnetic fields from a source can be amplified or enhanced in amplitudes during the transmission process and this contributes to the recovery of the high resolution but a fast decaying component of the image below the diffraction limit [2, 5, 7].

Following Cummer [22], we carried out a simple but robust full-wave FDTD simulation of a linear negative-index slab without including the often controversial issues of signs, series, and divergence, etc. which often lead to contradictory conclusions on the possibility of achieving perfect focusing [22]. The observed near-perfect focusing achieved in this study is linked to evanescent waves refocusing via well-defined resonant surface plasmons. This focusing effect is purely due to the material properties of the superlens used in our simulations.

We extended the idea of single-interface resonance [9] or surface plasmons resonance [2, 3] to evanescent wave-refocusing [6, 7]. This resonance phenomenon was first proposed by Pendry [2], and it occurs when both the transmission and reflection coefficients become divergent through the individual interfaces during the transmission process. Two mechanisms have been proposed to amplify the evanescent waves: single-interface resonance and overall resonance [9]. The overall resonance essentially involves direct divergence during the transmission process. In the nutshell, the physical consequences of these resonance phenomena when they occur is that wave energy can be seen to be pumped into the transmitted and reflected fields [9].To amplify the evanescent waves inside the superlens, an operation closer to exact resonance is required, and this can be achieved easily by single-interface resonance [9].

Instead of exploiting the divergence of the transmitted and reflected coefficients to excite the single interface resonance, we used a subtle but nuanced approach to excite it. We also examined the interplay closely between the amplified fields and the resonance [9] by monitoring the resonant coupling of the diverging cylindrical transmitted beams to the converging refracted beams at the second interface. This coupled interaction led to the excitation of a well-defined pair of resonant surface plasmons and their amplitudes increase exponentially as they propagate along with the interface and move in a direction that is opposite to each other. The coupling of the pair of resonant surface plasmons at the center of the interface led to the excitation of single-interface resonance or surface plasmons' resonance predicted by Pendry in [2]. Consequently, this resonance phenomenon contributes to the amplification of fields inside the slab and the refocusing of the evanescent near-fields in the image.

## II. THEORETICAL FORMULATIONS OF THE 2D LORENTZ MODEL

Imagine a harmonic oscillator characterized by a resonant frequency such that when a small driving force is applied to it, a very large displacement can be generated. We can model material as a collection of harmonically bound charges and the negative resonance of this material can be linked to a negative response experienced by this material. While the electric field, acting on the bound charges corresponds to the force, the response of this material, which is the dipole moment, corresponds to the displacement. The resonance excited in this material leads to negative values for or above the resonance [5]. Let us consider the response of the harmonic oscillator expressed in frequency domain [24].

$$P_i(\omega) = \varepsilon_0 \frac{\omega_{pe}^2}{\omega_{0e}^2 - \omega^2 + j\Gamma_e \omega} E(\omega) \tag{1}$$

Assuming charges are moving in the same direction as that of electric fields. For small losses $\frac{\Gamma_e}{\omega_0} \ll 1$, the response of this system becomes resonant at a characteristic frequency $f_0$. Using Lorentz model to study the temporary response associated with a component of the polarization $P_i(\omega)$ and its corresponding component of the electric field, $E(\omega)$ we have:

$$\frac{d^2}{dt^2} P_i(\omega) + \Gamma_e \frac{d}{dt} P_i(\omega) + \omega_{0e}^2 P_i(\omega) = \varepsilon_o \omega_{pe}^2 E(\omega) \tag{2}$$

Where, $\Gamma_e$ is the damping coefficient, $\omega_{0e}$ is the resonance frequency, $\omega_{pe}$ is the plasma frequency. From the left-hand side of Eq. (2), the first term accounts for the acceleration of charges; the second term accounts for the damping mechanism of the system; while the third term accounts for the restoring force with a characteristics frequency $\omega_0$.

Permittivity and permeability can be used to describe a negative-index material [1] and considering the constitutive relations of an isotropic, homogenous and dispersive Lorentz model which are given by [26, 30].

$$D(\omega) = \varepsilon(\omega) E(\omega) = \varepsilon_0 [\varepsilon_\infty + \chi_e(\omega)] E(\omega) \tag{3}$$

$$B(\omega) = \mu(\omega) H(\omega) = \mu_0 [\mu_\infty + \chi_m(\omega)] H(\omega) \tag{4}$$

Where $\mu_\infty$, $\varepsilon_\infty$ are the values of the permittivity and permeability at very high frequencies or optical frequencies, $\varepsilon(\omega)$, and $\mu(\omega)$ correspond to the complex permittivity and permeability respectively, $\chi_e(\omega)$ and $\chi_m(\omega)$ are the electric and magnetic susceptibilities. The Lorentz model describes the frequency-dependent behavior of the constitutive parameters in Eqs. (3) and (4) [26] and was used to simulate the electromagnetic scattering effects of lossy and lossless negative-index slabs in this study.

Where $\chi_e(\omega) = \dfrac{\omega_{pe}^2}{\omega_{0e}^2 - \omega^2 + j\Gamma_e \omega} \tag{5}$

And $\chi_m(\omega) = \dfrac{\omega_{pm}^2}{\omega_{0m}^2 - \omega^2 + j\Gamma_m\omega}$ (6)

$\omega_{pe}, \omega_{pm}$, are the plasma frequencies and $\omega_{0e}, \omega_{0m}$ are the resonance frequencies.

Substitute Eq. (5) in Eq. (3):

$$D(\omega) = \varepsilon_0 \varepsilon_\infty E(\omega) + \varepsilon_0 \left( \dfrac{\omega_{pe}^2}{\omega_{0e}^2 - \omega^2 + j\Gamma_e\omega} \right) E(\omega) \quad (7)$$

Let's assume an auxiliary term

$$I_k(\omega) = \varepsilon_0 \left( \dfrac{\omega_{pe}^2}{\omega_{0e}^2 - \omega^2 + j\Gamma_e\omega} \right) E(\omega) \quad (8)$$

Eq. (8) in Eq. (7) result in

$$D(\omega) = \varepsilon_0 \varepsilon_\infty E(\omega) + I_k(\omega) \quad (9)$$

Performing some simple arithmetic with Eq. (8) leads to:

$$-\omega^2 I_k(\omega) + j\omega\Gamma_e I_k(\omega) + \omega_{0e}^2 I_k(\omega) = \varepsilon_0 \omega_{pe}^2 E(\omega) \quad (10)$$

Repeating the procedures above using Eq. (4) and Eq. (6) and assuming an auxiliary term $G_k(\omega)$:

$$B(\omega) = \mu_0 \mu_\infty H(\omega) + \mu_0 \left( \dfrac{\omega_{pm}^2}{\omega_{0m}^2 - \omega^2 + j\Gamma_m\omega} \right) H(\omega) \quad (11)$$

$$G_k(\omega) = \mu_0 \left( \dfrac{\omega_{pm}^2}{\omega_{0m}^2 - \omega^2 + j\Gamma_m\omega} \right) H(\omega) \quad (12)$$

Substituting Eq. (12) in Eq. (11):

$$B(\omega) = \mu_0 \mu_\infty H(\omega) + G_k(\omega)) \quad (13)$$

From Eq. (12):

$$-\omega^2 G_k(\omega) + j\omega\Gamma_m G_k(\omega) + \omega_{0m}^2 G_k(\omega) = \mu_0 \omega_{pm}^2 H(\omega) \quad (14)$$

Transform Eqs. (10) and (14) to time domain using Auxiliary Differential Equation (ADE) method [26-30].

$$\frac{d^2}{dt^2}I_k(t)+\Gamma_e\frac{d}{dt}I_k(k)+\omega_{0e}^2 I_k(t)=\varepsilon_0\omega_{pe}^2 E(t) \tag{15}$$

$$\frac{d^2}{dt^2}G_k(t)+\Gamma_m\frac{d}{dt}G_k(k)+\omega_{0m}^2 G_k(t)=\mu_0\omega_{pm}^2 H(t) \tag{16}$$

Eq. (15) and Eq. (16) are suitable for discretization in the FDTD space.

At optical frequencies, the three terms in Eq. (2) reduce to the free space. When the Lorentz model becomes resonant, both the real parts of the permittivity and permeability become simultaneously negative in a narrow frequency range: ($\omega_{0e,0m}, \sqrt{\omega_{0e,0m}^2+\omega_{pe,pm}^2}$), just above the resonance [24, 30].

For implementation in the FDTD simulator, we transform Eqs. (9), (13), (15), and (16) to discrete time domain. This may allow temporary response of the oscillator to be obtained using inverse Fourier transformation and the two classical Maxwell's curl equations.

### III. 2D FINITE DIFFERENCE TIME DOMAIN ALGORITHM

The 2D FDTD algorithm solves the electromagnetic problem of TM sets for $H_x$, $H_y$ and $E_z$. Both the electric and magnetic fields were discretized in space and time. In other words, these fields were allocated in space and matching in time for the evolution of the procedure using Yee's leap-frog algorithm [31]. This robust scheme considers the electric and magnetic fields shifted in space by half a cell and in time by half a time step using the central difference approximations of their derivatives. The discretized equations for the Lorentz- metamaterial model are given below:

$$I_k^{n+1}=\left[\frac{2-\Delta t^2\omega_{oe}^2}{1+0.5\Delta t\Gamma_e}\right]I_k^n+\left[\frac{0.5\Delta t\Gamma_e-1}{0.5\Delta t\Gamma_e+1}\right]I_k^{n-1}+\left[\frac{\Delta t^2\varepsilon_0\omega_{pe}^2}{1+0.5\Delta t\Gamma_e}\right]E_k^n \tag{17}$$

$$E_z^{n+1}(i,j)=\left(\frac{2\varepsilon-\Gamma_e\Delta t}{2\varepsilon+\Gamma_e\Delta t}\right)E_z^n(i,j)+\frac{2\Delta t}{2\varepsilon+\Gamma_e\Delta t}\left[\frac{H_y^n(i,j)-H_y^n(i-1,j)}{\Delta x}\right]$$
$$-\frac{2\Delta t}{2\varepsilon+\Gamma_e\Delta t}\left[\frac{H_x^n(i,j)-H_x^n(i,j-1)}{\Delta y}\right] \tag{18}$$

$$D_z^{n+1}(i,j)=D_z^n(i,j)+\frac{\Delta t}{\Delta x}\left(H_y(i,j)-H_y(i-1,j)\right)-\frac{\Delta t}{\Delta y}\left(H_x(i,j)-H_x(i,j-1)\right) \tag{19}$$

$$E_z^n(i,j)=\frac{(D_z^n-I_k^n)}{(\varepsilon_\infty\varepsilon_0)} \tag{20}$$

$$G_k^{n+1} = \left[\frac{2 - \Delta t^2 \omega_{om}^2}{1 + 0.5\Delta t \Gamma_m}\right] G_k^n + \left[\frac{0.5\Delta t \Gamma_m - 1}{0.5\Delta t \Gamma_m + 1}\right] G_k^{n-1} + \left[\frac{\Delta t^2 \mu_0 \omega_{pm}^2}{1 + 0.5\Delta t \Gamma_m}\right] H_k^n \tag{21}$$

$$B_y^{n+1}(i,j) = B_y^n(i,j) + \frac{\Delta t}{\Delta x}\left(E_z^n(i+1,j) - E_z^n(i,j)\right) \tag{22}$$

$$B_x^{n+1}(i,j) = B_x^n(i,j) - \frac{\Delta t}{\Delta y}\left(E_z^n(i,j+1) - E_z^n(i,j)\right) \tag{23}$$

$$H_z^n(i,j) = \frac{(B_y^n - G_k^n)}{(\mu_\infty \mu_0)} \tag{24}$$

$$H_x^{n+1}(i,j) = \left(\frac{2\mu - \Delta t \Gamma_m}{2\mu + \Delta t \Gamma_m}\right) H_x^n(i,j) - \frac{2\Delta t}{2\mu + \Delta t \Gamma_m}\left[\frac{E_z^n(i,j) - E_z^n(i,j-1)}{\Delta y}\right] \tag{25}$$

$$H_y^{n+1}(i,j) = \left(\frac{2\mu - \Delta t \Gamma_m}{2\mu + \Delta t \Gamma_m}\right) H_y^n(i,j) - \frac{2\Delta t}{2\mu + \Delta t \Gamma_m}\left[\frac{E_z^n(i+1,j) - E_z^n(i,j)}{\Delta x}\right] \tag{26}$$

The procedures for implementation in the FDTD algorithm are given below:

(I) Update $I_k$ in Eq. (17) using the previous values of $E$ and $I_k$.

(II) Update $E$ in Eq. (18) in free space and add a source.

(III) Update $D$ in Eq. (19) in the NEGATIVE-INDEX slab values of $D$ and $H$.

(IV) Update $E$ in Eq. (20) using the previous values of $D$ and $I_k$.

(V) Apply the absorbing boundary conditions at the termination points.

(VI) Repeat the above procedures for the magnetic fields using Eq. (21) – Eq. (26).

## IV. 2D-FDTD SIMULATION RESULTS AND DISCUSSIONS

The XY-FDTD simulation domain was 400 cells in the x-direction and 400 cells in the y-direction for the lossy and lossless negative index slabs. The cell sizes are given by $\Delta x = \Delta y = \lambda/40$.

The time step is given:

$$\Delta t = \frac{0.95}{c_0 * \sqrt{\left(\frac{1}{\Delta x}\right)^2 + \left(\frac{1}{\Delta y}\right)^2}} = 2.239 \times 10^{11} s$$

The FDTD grid was terminated by MUR, ABC on all the four sides of the simulation domain [32]. A line source was positioned at a distance of $0.5\lambda$ from the slab with thickness $d = 2\lambda$. The slab was matched to free space such that $w_{0e} = w_{0m} = w_0$, $w_{pe} = w_{pm} = w_p$ and $\Gamma_e = \Gamma_m = \Gamma$. A continuous wave pulse was used to investigate the negative refraction of the slab. The center frequency of the line source was 6 MHz and was driven by a continuous wave pulse [28. 33].

To visualize the negative refraction, we use a lossless slab with the following parameters: $w_0 = 9.9965 \times 10^8 \, rad/s$, $w_p = 6.9285 \times 10^9 \, rad/s$ and $\Gamma = 0.0$. The slab was located at the grid points, X = (20, 380) and Y = (120, 200) in the simulation domain and has a total thickness of $2\lambda$. Negative refraction can be clearly observed in Fig. 1 when the beam interacts with the slab after a few cycles. Backward propagating waves are clearly visible and this is consistent with previous observations [34].

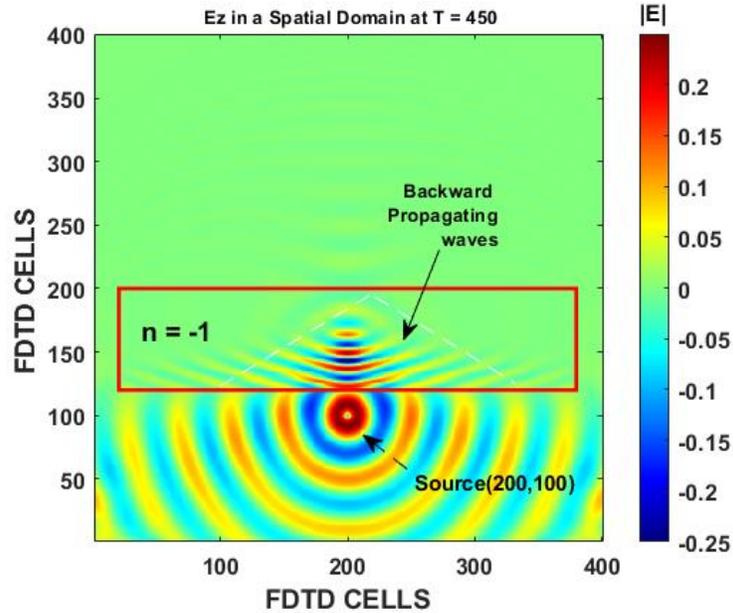

FIG. 1. The electric field distribution over the FDTD domain at $t = 450\Delta t$ for a lossless slab.

**V. EXCITATION OF RESONANT SURFACE PLASMONS**

Let us revisit the condition for exciting the surface plasmons at the interface of a negative-index medium [35].

$$\frac{k_z^{(1)}}{\varepsilon_1} + \frac{k_z^2}{\varepsilon_2} = 0 \tag{27}$$

This results in the dispersion equation:

$$k_x = \frac{\omega}{c}\left[\frac{\varepsilon_2(\varepsilon_2 - \mu_2)}{\varepsilon_2^2 - 1}\right]^{1/2} \qquad (28)$$

Assuming that $\varepsilon_1 = 1$ and $\mu_1 = 1$ for a positive index medium. The condition for Eq. (27) can be satisfied for the imaginary wave vector $k_z^{(2)}$ only when $\varepsilon_2$ is negative. The causal plasma dispersion forms $\varepsilon_2 = 1 - \frac{\omega_p^2}{\omega^2}$, $\mu_2 = 1 - \frac{\omega_{mp}^2}{\omega^2}$ are assumed for the negative index material [36, 37, and 38]. For a disperless and lossless slab such that $\varepsilon_1 = \varepsilon_3, \mu_1 = \mu_3 = 1, \varepsilon_2^1 = 0$, two degenerate surface plasmons get coupled at the two interfaces. This leads to the excitation of coupled slab modes: a symmetric and anti-symmetric modes and the resonance condition for the surface plasmons are given as [35]:

$$\tanh(k_z^{(2)} d/2) = -\varepsilon_2 k_z^{(1)} / k_z^{(2)} \qquad (30)$$

$$\coth(k_z^{(2)} d/2) = -\varepsilon_2 k_z^{(1)} / k_z^{(2)} \qquad (31)$$

Considering the anatomy of a superlens [3], there exists is a subtle contribution of the resonant surface plasmons to the amplification of near-fields and to the refocusing of these fields in the image [2, 7]. Let us imagine the wave-fields of the object, the anti-surface plasmons wave-fields, and the surface plasmons wave-fields in a silver slab [3]. Matching these fields selectively at the boundary of the slab can lead to the excitation of the resonant surface plasmons on the second interface of the slab [3]. These resonant surface plasmons can invariably amplify the nonpropagating near-fields across the slab and restore them in the image plane with their correct amplitudes [5]. The interplay between surface plasmons and the amplification of the near-fields was initially predicted by Pendry in [2] and later investigated by Ruppin in [39]. We investigated these surface plasmons by closely monitoring the coupled interaction of the transmitted cylindrical beams and the converging refracted beams. We found that for every transmitted cylindrical beam interacting with a converging refracted beam at the second interface, a pair of resonant surface plasmons is excited as shown in Figs. 3(a-c). These modes propagate along with the air-slab interface with amplitudes that increase exponentially as shown in Fig. 3(b).In an attempt to visualize these resonances [40], FDTD simulation was performed using a negative-index slab with a refractive index, $n \approx -1$. The thickness the slab was $d = 2\lambda$ while the source-to-slab distance of $0.5\lambda$ was used. A Gaussian beam with waist $0.9\lambda$ was launched toward the slab. The following parameters: $w_0 = 9.9965 \times 10^8 \, rad/s$, $w_p = 6.9285 \times 10^9 \, rad/s$, $\Gamma = 1 \times 10^7 \, rad/s$ were used in the simulated results in Figs. (3) (a-c).

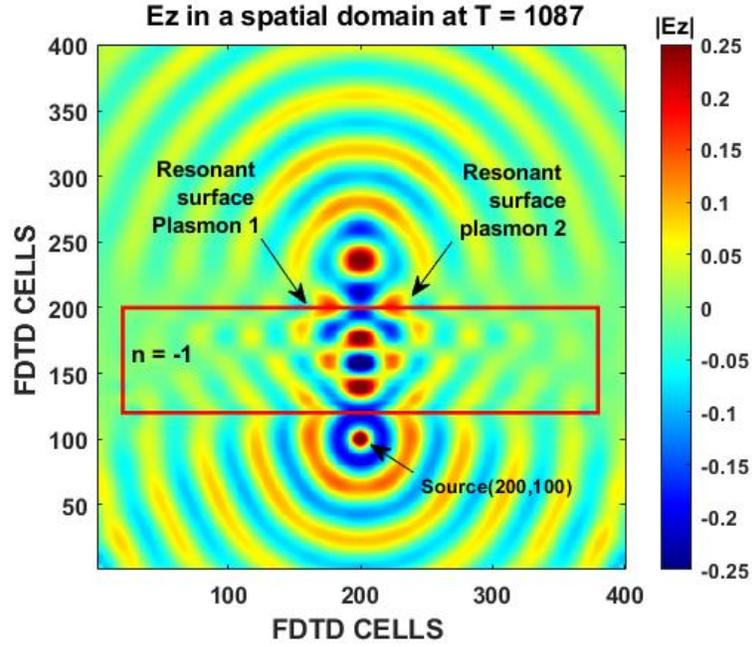

FIG. 3. (a) The electric field distribution over the FDTD simulation domain at t = 1087 $\Delta t$ for a lossy slab.

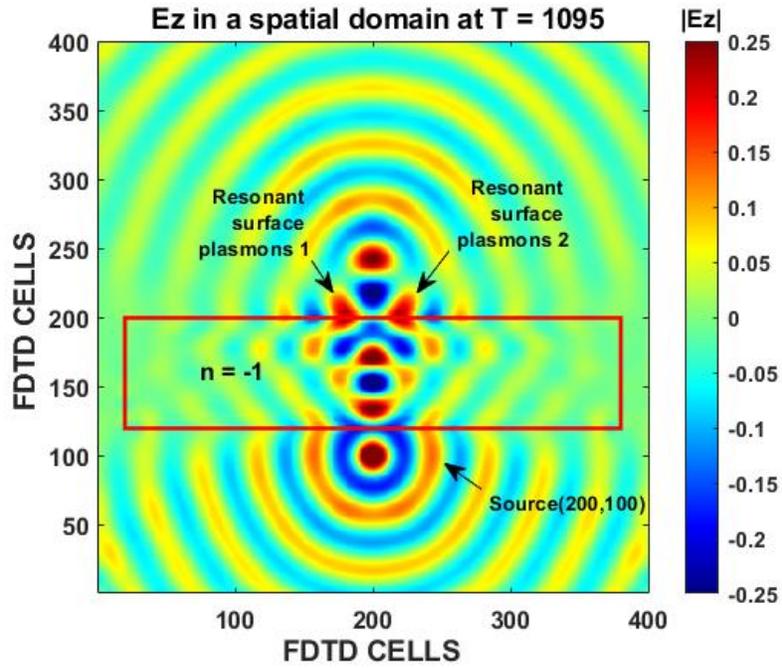

FIG. 3. (b) The electric field distribution over the FDTD simulation domain at t = 1095 $\Delta t$ for a lossy slab.

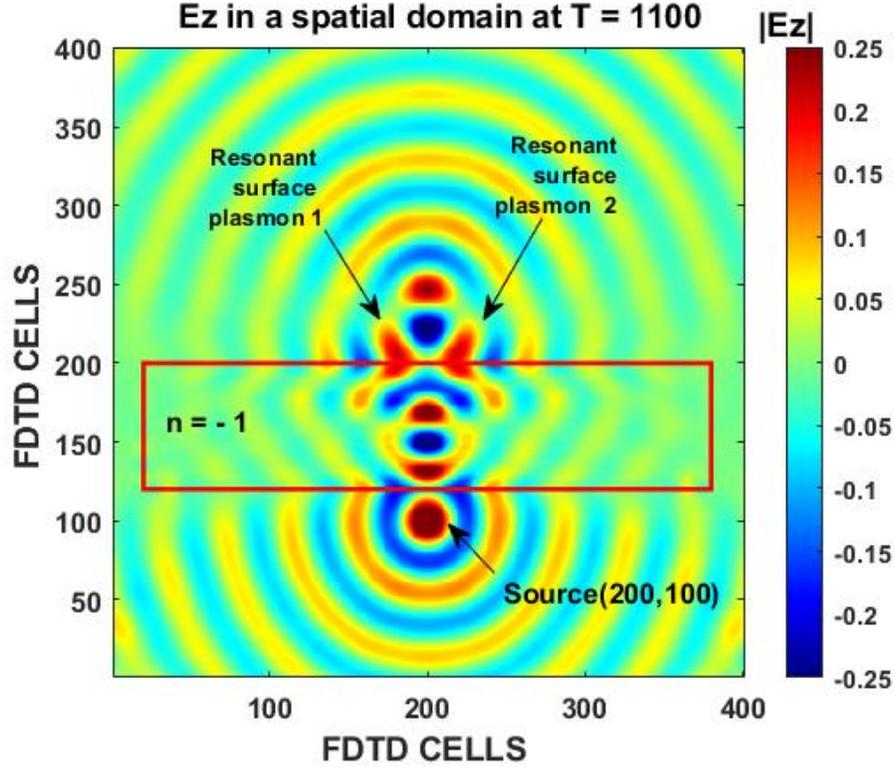

FIG. 3. (c) The electric field distribution over the FDTD simulation domain at t = 1100 $\Delta t$ for a lossy slab.

In Figs. 3(a-c), the interaction of the beam with the first interface of the slab results in negative refraction. The refracted waves inside the slab gradually converge towards the center of the slab and subsequently get coupled to the divergent transmitted cylindrical beam at the second interface. This coupled interaction led to the excitation of a pair of resonant surface plasmons which propagate in opposite direction to each other and with amplitudes that increase exponentially as they move toward the center of the slab.

A resonant surface plasmon can be observed to propagate from the right-hand side of the slab towards its center with amplitude that increases exponentially as they propagate along with the interface. Conversely, another resonant surface plasmon also propagates from the left-hand side of the slab towards the center with amplitude that also increases exponentially as they propagate along with the interface. These modes have identical amplitudes and propagate with similar velocities. The tail ends of these modes release wave energy into the slab and beyond. This partly contributes to the growing exponential fields inside the slab as well as refocusing of the evanescent near-fields as shown in Figs. 3(a, c).

It is interesting to know that, the resonant interaction of these modes leads to a resonance phenomenon at the second interface of the negative-index slab in Fig. 4(a). This may be called single-interface resonance. This resonance phenomenon was first predicted by Pendry in [3] and subsequently investigated by Luo et al. in [9]. It occurs when both transmission and reflection coefficients becoming divergent during the transmission process and wave energy is pumped into the reflected and transmitted fields [9]. However, in this study, the single-interface resonance was excited by the coupling of the resonant surface plasmons as demonstrated in Fig. 4(a). Besides, incident wave energy can be seen to be pumped into the slab and beyond as demonstrated in Figs. 4(b) and 4(c). A part of this wave energy is transmitted to the image plane and contributes immensely to the refocusing of the near-fields in the image. While the other part of the wave is reflected back into the slab and supports the resonant amplification of fields inside the slab. These amplified fields are otherwise known as growing exponential effects [25, 41]. These resonance phenomena lead to the amplification and the recovery of some evanescent near-fields in the image as demonstrated by simulated results in Fig. (3) and Fig. (4).

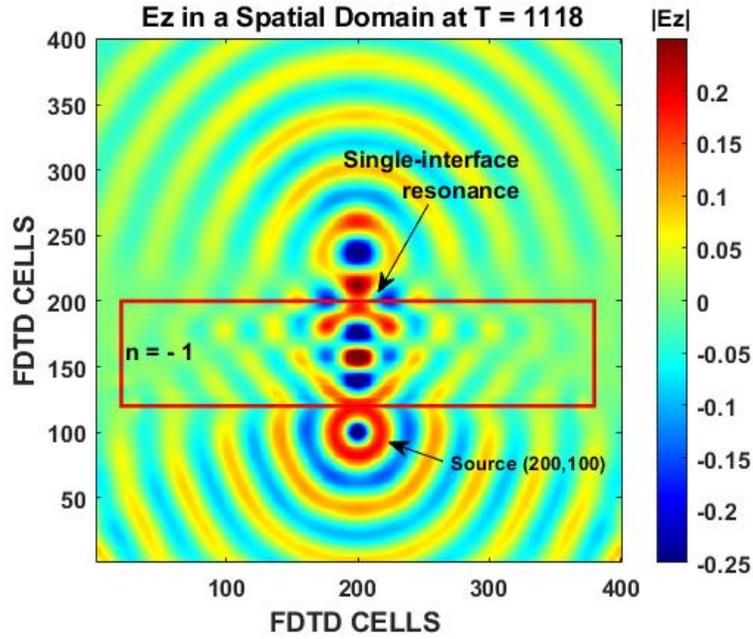

FIG. 4. (a) The electric field distribution over the FDTD domain at t = 1115 $\Delta t$ for a lossy negative-index slab.

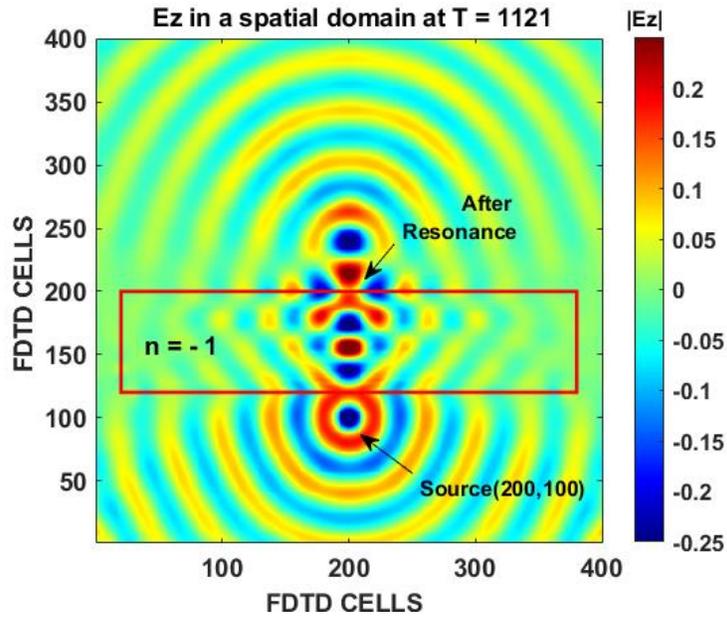

FIG. 4. (b) The electric field distribution over the FDTD domain at t = 1118 $\Delta t$ for a lossy negative-index slab.

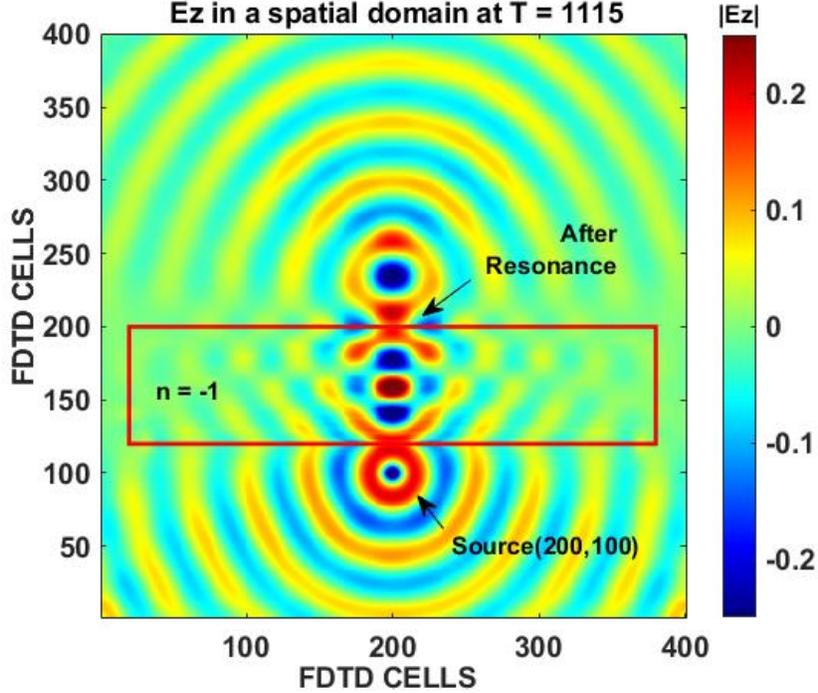

FIG. 4. (c) The electric field distribution over the FDTD domain at t = 1125 $\Delta t$ for a negative-index slab.

## VI. SUPERLENSING EFFECTS

The idea of capturing the high-resolution but rapidly decaying components of the image is rooted in the process of resonant amplification of the fields [8]. In addition to that; a superlens is predicted to significantly amplify the evanescent near-fields by compensating for their loss outside of it and restoring them in the image [5, 6]. This enables high-resolution imaging with sub-wavelength features of the object to be achieved beyond the limit of diffraction [2]. Sub-wavelength imaging by a superlens has been linked to the transmission and recovery of the nonpropagating near-fields through it [5-7, 9]. This transmission used in this context is fundamentally different from the conventional one that involves transportation of energy [9] and the nonpropagating fields do not transport energy since they decay exponentially in the direction of wave propagation [2]. The transmission of these near-fields was aided by the resonance phenomenon demonstrated in Figs. (3) and (4) above. The superlens used here consists of a negative-index slab embedded in a positive index medium.

In an attempt to investigate this superlensing effect in a lossless/lossy planar negative-index slab, FDTD simulation was carried out. This flat superlens is characterized by a refractive index, n $\approx$ -1, $\mu$ = -1, $\varepsilon$ = -1, with a thickness $d = 2\lambda$. The source-to-slab distance of $0.5\lambda$ was used during the simulations. And a line source with a center frequency of 6-MHz and waist $0.9\lambda$ was driven into the slab using continuous wave pulses. After few circles, superlensing effect could be observed in Fig. 5(a) which shows the electric field distribution generated by the point source. The beam was negatively refracted upon reaching the first interface of the superlens. Growing exponential fields can also be observed inside the superlens and this has been reported in [23]. These observations are consistent with those obtained in [23]. At the second interface, the refracted beam coupled to the diverging transmitted beam to form a pair of resonant surface plasmons. These modes transport part of the incident wave energy along the interface. Consequently, the resonance interaction of these modes results in the refocusing of the evanescent near-fields in the image plane in Fig. (4) and Fig. (5).

## VII. THE EFFECT OF MATERIAL LOSS

We went further to investigate the effects of material loss on the fine subwavelength features of the image. The simulated results are shown in Figs. (5), (6), and (7) for different losses. Fig. (5) depicts the FDTD result for a lossless superlens, $\Gamma = 0.0$ while simulated results in Fig. (6) and Fig. (7) are characterized by material losses: $\Gamma = 1\times10^8 \, rad/s$ and $\Gamma = 1\times10^7 \, rad/s$ respectively.

Following Luo et al. in [14], we interpreted the image of the point source as an intensity peak in Fig. (5), with a resolution of $\sim 0.6\lambda$. This was measured by the considering the distance between the intensity peak and the nearest minima. The transverse size of this image was $\sim 0.9\lambda$ measured along the y-axis and this is always limited by a fraction of a wavelength [9].

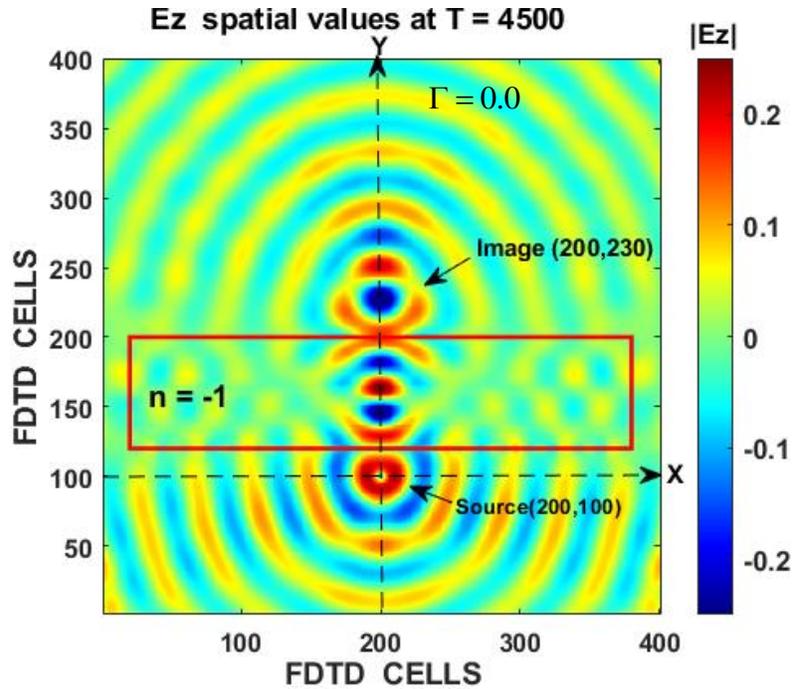

FIG. 5. The electric field $E_z$ of a point source and its image across the superlens at t = 4500 $\Delta t$.

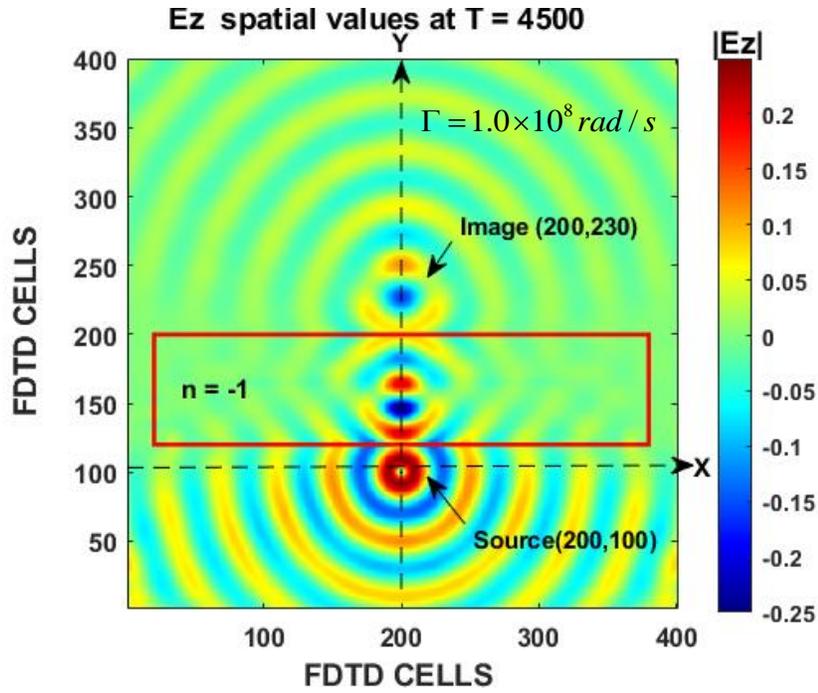

FIG. 6. The electric field $E_z$ of a point source and its image across the superlens at t = 4500 $\Delta t$ .

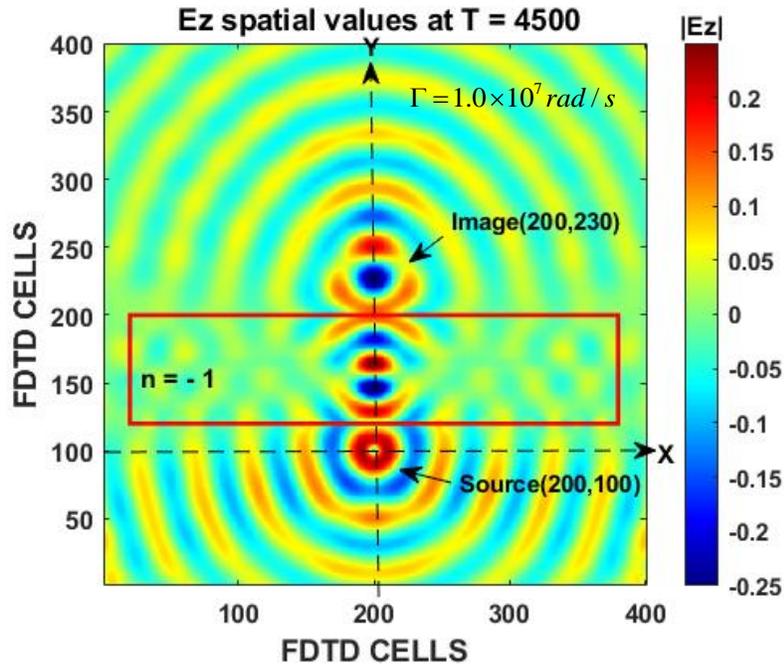

FIG. 7. The electric field $E_z$ of a point source and its image across the superlens at t = 4500 $\Delta t$

The effects of the material loss on the amplitudes of the electric fields recorded at the image location were quantified in Fig. (8). A receiver was placed at the image location to detect and record the image fields during the simulation for different losses. The spectra of the recorded electric fields for different losses $\Gamma = 0.0, \Gamma = 1\times 10^8 \, rad/s$, $\Gamma = 1\times 10^7 \, rad/s$ are shown in Fig. (8). The lost in the amplitude of the fields recorded is very apparent in Fig. (6), and sub-wavelength features that disappeared in this figure are due to the loss in the amplitudes of the image fields for the material loss, $\Gamma = 1\times 10^8 \, rad/s$. From Fig. (8), the sharp decrease in the amplitude of the image field at high frequencies can be linked to the loss of the sub-wavelength features in Fig. (6). The strength of growing exponential fields is attenuated inside the superlens. Apparently, the fine-detailed sub-wavelength features in the central peak of the image begin to disappear eventually (Fig. (6)). To minimize the effect of material loss on the resolution of the image, a slightly low lost value of $\Gamma = 1\times 10^7 \, rad/s$ was used. The subwavelength features are optimized when we compare Fig. (5) and Fig. (7). This can also be observed in Fig. (8).

## IX. THE QUALITY OF THE FOCUSED IMAGE

We verify the quality of the focusing by the superlens by comparing the spectra of the electric fields recorded at both source and image positions. The results are shown in Fig. (9). The half-width of the primary peak; 0.03GHz of the source field was $25\lambda$, while the half-width of the predicted peak of 0.025GHz of image field was $30\lambda$. These two estimates show how the superlens focuses the source field nicely beyond the planar superlens.

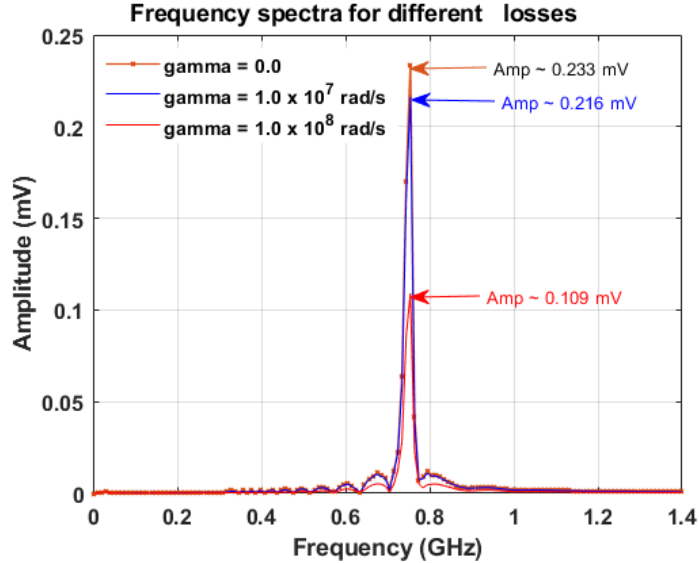

FIG. 8. The spectra of electric fields $E_z$ recorded at the image position for different material losses.

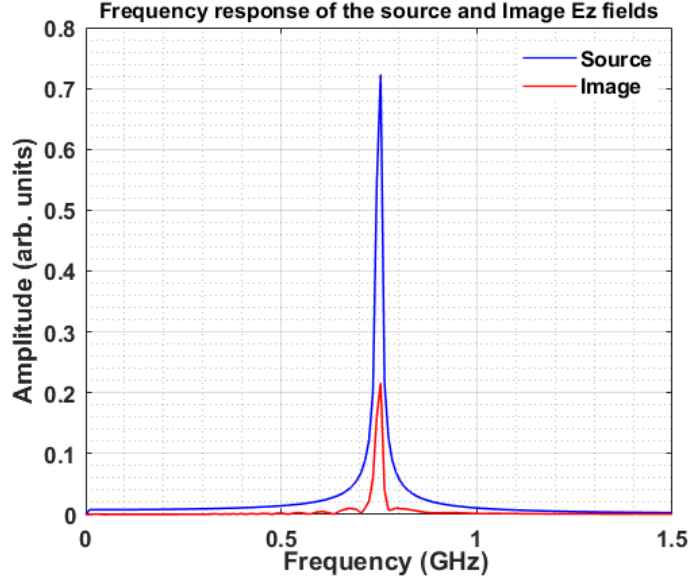

FIG. 9. The Spectra of the Image and Source Fields

The simulated results above clearly demonstrated the numerical evidence of the superlensing effect of a planar negative –index slab (superlens). There is also apparent distortion in the image formed by the superlens which may be attributed to the surface plasmons resonances. These resonances have been predicted to affect the image disproportionately and must be included in practical imaging applications [8]. To minimize these effects, we optimized the thickness of the superlens in the following section. We further explore the contribution of the thickness of the superlens to the refocusing of the near-fields and reconstruction of sharp image in the section that follows.

## X. SUBWAVELENGTH FOCUSING AND DOUBLE-FOCUSING EFFECT

Pendry [2] initially pointed out a subtle link exists between the focusing capability of a superlens and the excitation of well-defined surface plasmons. Following this suggestion, Fang et al. [7] carried out the analysis of the evanescent wave-refocusing via the excitation of surface plasmons [6]. Using a thin-film of silver as a superlens, the mechanism mentioned above was investigated with respect to optical lithography. In this experiment, the superlens was used to transfer the image of a pattern (NANO) lithographically written to a photoresist by coaxing an evanescent wave to grow inside of it. Using thicker silver superlens will cause the material loss to dominate the evanescent wave refocusing [6] and this will affect the degree of subwavelength focusing that can be achieved by a superlens. Conversely, the thickness of the superlens was optimized in this experiment such that evanescent wave-refocusing dominates the material loss and this paves a way to the reconstruction of the optical image of the object with sub-wavelength features [6, 7].

Subwavelength focusing involves the recovery of the evanescent waves in the image [42] and this has been investigated previously by several authors using photonic crystals [12, 13, and 14] and metamaterials [7, 15-21]. A perfect lens is capable of focussing all the Fourier components of the source fields in the image which also includes the evanescent components [2].In an attempt to observe perfect focusing by the negative-index slab, FDTD numerical simulation was carried out by Cummer in [22]. The leap-frog numerical scheme was used in this study to simulate electromagnetic interactions with the slab with the aim of observing the perfect focusing. Even though perfect-focusing was not achieved, but a subwavelength focusing was achieved which involves the recovery of some evanescent waves in the image. We revisited this perfect focusing concept to gain deep insights into some of the trivial mechanisms or wave processes that may affect it besides the dominant effect of the material loss [2, 4]. We also extended our results to the evanescent wave refocusing [6, 7] and optimization of the superlens thickness in order to observe the double-focusing effect in the lens. The advantage of a superlens over the conventional lenses is that a complete image of the object can be obtained which may find wide applications in biological and fluorescence imaging [4]. In general, superlens can also serve as a near-perfect lens with its resolution limited only by the absorption in the lens [4].

In an attempt to view this double-focusing effect in the superlens [1, 2, and 15], we carried out the FDTD simulations on a lossy slab with the following parameters: $w_0 = 9.9965 \times 10^8 \, rad/s$, $w_p = 6.9285 \times 10^9 \, rad/s$, $\Gamma = 1 \times 10^7 \, rad/s$. The source-to-slab distance was kept at $|y_0| = 0.5\lambda$. The line source with a center frequency of 6-MHz was located at 20-cells in front of the slab in anticipation of observing the double-focusing effect. The simulated results for different thicknesses of the slab, d = $0.23\lambda$, $0.5\lambda$ $1.20\lambda$, and $1.43\lambda$ are shown in Figs. (10) and (11) respectively. It is apparent that for thickness less than or equal to the source-to-slab distance, there is no coupling of the diverging transmitted and refracted beams. By extension resonant surface plasmons were not excited. Consequently, no visible image was formed in the simulated results in Figs. (10). However, when the thickness is twice or more greater than the source-to-superlens distance, the image of the source is clearly visible is Figs. 11(a, b)).

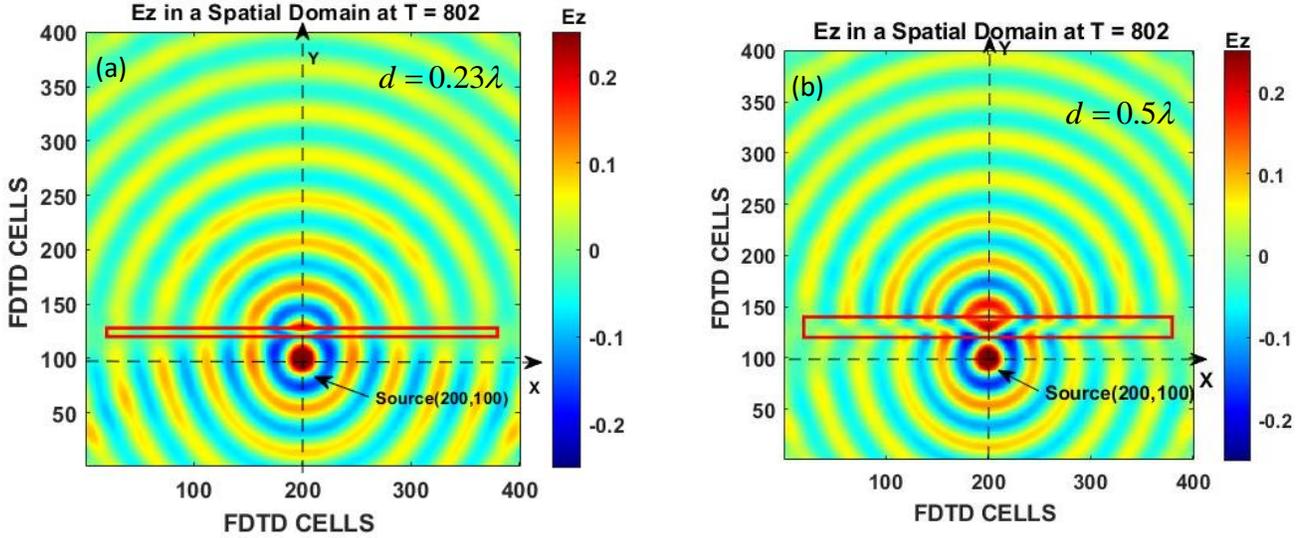

FIG. 10. (a) The electric field distribution over the FDTD domain at $t = 802\Delta t$ for a negative-index slab with a thickness $d = 0.23\lambda$. (b) The electric field distribution over the FDTD domain at $t = 802\Delta t$ for a negative-index slab with a thickness $d = 0.50\lambda$.

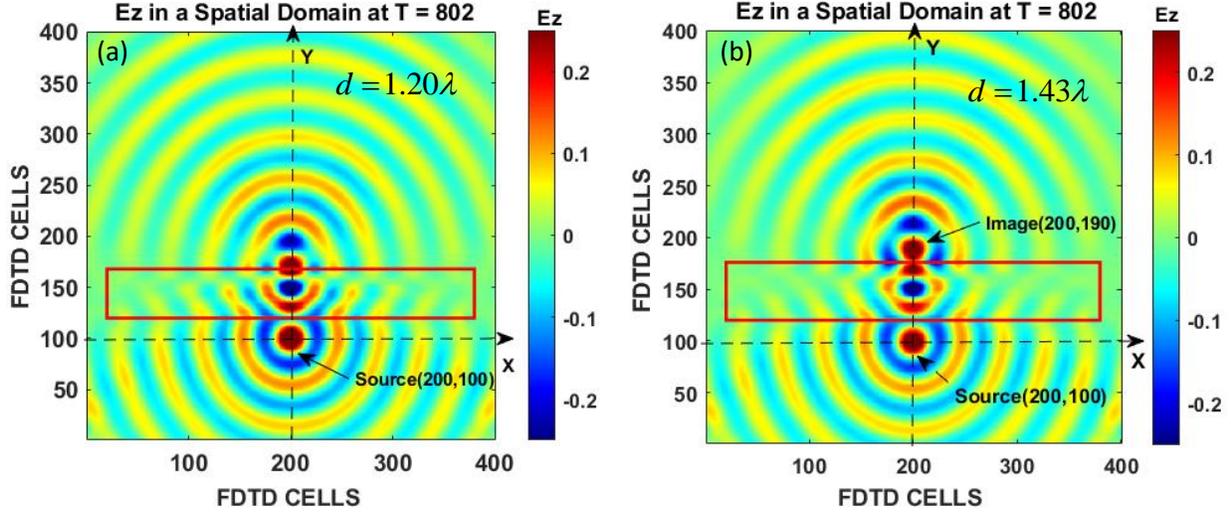

FIG. 11. (a) The electric field $E_z$ of a point source and its image across the superlens at $t = 802\Delta t$ with a thickness $d = 1.2\lambda$. (b) The electric field $E_z$ of a point source and its image across the superlens at $t = 802\Delta t$ with a thickness $d = 1.43\lambda$.

The image of the object was focussed right at the interface of the superlens for $d = 1.20\lambda$ in Fig. 11(a). A similar observation was reported elsewhere by Grbic and Eleftheriades in [15, Fig. (13)]. When the thickness of the superlens was increased to $d = 1.43\lambda$, a clear and sharp image of the object becomes apparent as shown in Fig. 11(b) with a resolution of ~ $0.58\lambda$. The focal length of the image beyond the superlens in Fig. 11(b) is $0.43\lambda$ respectively. Comparing the simulated result in Fig. 11(b) to those obtained in Fig. (5) and Fig. (7), it is evident that the distortion in the image in Figs. (5) and (7) was significantly improved. In other words, the thickness of the superlens was optimized in Fig. 11(b) such that evanescent wave refocusing dominates the material loss, and the effect of the resonance was also minimized which led to intense focusing by the superlens. Putting it simply, we effectively minimize the resonant effect of the surface plasmons which led to the reconstruction of a high-resolution image of the source demonstrated in Fig. 11(b). Consequently, these resonant surface plasmons substantially contribute to the amplification and recovery of evanescent near-fields in the image. It is also quite remarkable to observe that the image of the source began to be appearing inside the superlens though with some aberrations as demonstrated in Fig. 11(b). These aberrations or distortions may be linked to the resonant effects inside the superlens during the transmission process.

We have prescribed numerical approximations that may be useful in practical near-fields imaging setup using flat superlens. These may also be extended to photo crystal superlens. Besides, these numerically simulated results may also serve as useful approximations to the degree of amplification that will be optimal to exciting well-defined resonant surface plasmons [2] and achieving subwavelength focusing in planar and photonic superlenses especially in experimental setups.

______________________________________________